# The Kinetics of Chirality Assignment in Catalytic Single-Walled Carbon Nanotube Growth

Ziwei Xu,[1] Tianying Yan,[2] Feng Ding[1,*]

[1]*Institute of Textiles and Clothing, Hong Kong Polytechnic University,*

*Hong Kong, Peoples Republic of China*

[2]*Institute of New Energy Material Chemistry, College of Chemistry, Nankai University,*

*Tianjin 300071, Peoples Republic of China*

Chirality-selected single-walled carbon nanotubes (SWCNTs) ensure a great potential of building ~ 1 nm sized electronics. However, the reliable method for chirality-selected SWCNT is still pending. Here we present a theoretical study on the SWCNT's chirality assignment and control during the catalytic growth. This study reveals that the chirality of a SWCNT is determined by the kinetic incorporation of the pentagon formation during SWCNT nucleation. Therefore, chirality is randomly assigned on a liquid catalyst surface. Furthermore, based on the understanding, two potential methods of synthesizing chirality-selected SWCNTs are proposed: (*i*) by using Ta, W, Re, Os, or their alloys as solid catalysts, and (*ii*) by changing the SWCNT's chirality frequently during growth.

[*]Corresponding author : feng.ding@polyu.edu.hk

Due to the numerous potential applications, great efforts, including the chirality-selected growth and the post-growth selection, have been dedicated for a reliable method of producing SWCNTs with desired chiralities. Although significant progresses have been achieved in post-growth selection, such as the selection of more than 10 different SWCNTs in high purity by DNA wrapping [1] or gel chromatography [2], they suffer the drawback for only selecting short SWCNTs with low yields and high expense. Therefore direct growth of chirality-selected SWCNTs is the most desired method for achieving this goal.

Over the last two decades, significant attempts have been made on the growth of chirality-selected SWCNTs, but the achievements were very limited [3-7]. Up to date, the best technique is only able to grow a single type of SWCNT (e. g., the (6,5) [5, 6]and (9,8) [8]) with ~ 50% selectivity, which is far from that required for high performance electronics production of > 99% selectivity. Such slow progress of direct growth is attributed to the lack of understanding on the SWCNT growth mechanism, especially on the mechanism of chirality assignment and control during SWCNT nucleation and growth.

Here we present a theoretical analysis on SWCNT nucleation. The current study demonstrates that the nucleation of a SWCNT is a kinetic process, with random assignment of the chirality on a liquid catalyst particle surface. Based on such insightful understanding, the difficulty that hinders the chirality selection during SWCNT growth is well explained, and potential routes toward the chirality-selected SWCNT growth were proposed.

According to the vapor-liquid-solid (VLS) growth mechanism, the scenery of catalytic SWCNT growth is that a SWCNT is attached to a sphere-like liquid catalyst particle with its open end [Fig. 1(a)]. In such a continuous model, the differences among SWCNTs of various chiralities can not be distinguished. However, at atomic level, the open end of a SWCNT is a carbon ring of either armchair (AC), zigzag (ZZ), or chiral [Fig. 1(b-e, f-i)], through which the chirality of a SWCNT can be recognized by the catalyst [9].

It was widely argued that a SWCNT can be nucleated in a larger number if it has a more stable interface with the catalyst [7, 10, 11]. The validity of such mechanism

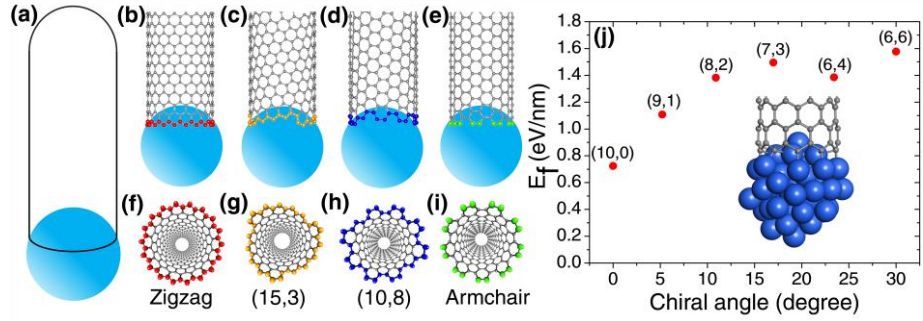

**FIG. 1** (color online). The formation of single-walled carbon nanotubes (SWCNT)-catalyst interface. (a) The vapor-liquid-solid (VLS) model of SWCNT growth. (b-e) Atomic details of (18,0), (15,3), (10,8), and (9,9) SWCNTs on the liquid catalyst particles. (f-i) Bottom view of the tubes end. (j) The interfacial formation energies (IFEs) between a series of SWCNTs on the liquid-like $Ni_{55}$ catalyst particle.

of chirality selectivity is expected, as each type of SWCNT has a unique open end attached to the catalyst surface [Fig. 1(b-e)].

Figure 1(j) shows the interfacial formation energies (IFEs) of a series of SWCNTs, with very similar diameters, on the liquid $Ni_{55}$ catalyst particle, calculated by the density functional theory (DFT) method incorporated in the VASP (Vienna Ab-initio Simulation Package) [12, 13]. The generized gradient approximation (GGA) with the Perdew-Burke-Ernzerhof (PBE) functions was employed during the calcuations [14, 15]. As expected, a systematic variation of the IFE, with the smaller the chiral angle the lower the IFE, can be clearly seen. This indicates that a small chiral angle SWCNT is more stable than a large chiral angle one. Assuming that the nucleation of a SWCNT is a *quasi-thermal-equilibrium* process, the number of each type of SWCNT can be estimated by

$$N \sim \mathrm{Exp}(-E_f/k_bT), \qquad (1)$$

where $E_f$ is the IFE of the SWCNT on the catalyst surface, $k_b$ is the Boltzmann constant, and $T$ is the typical experimental temperature of SWCNT growth, which are mostly in the range of 1,000 – 1,300K. As shown in Fig. 1(j), the difference in IFE varies in a range of 0.7--1.6 eV/nm. For a typical SWCNT of diameter $d \sim 1$ nm, the maximum IFE difference is $\Delta E_f \sim 0.9$ eV/nm * ($\pi * d$) ~ 3.0 eV. Thus, the maximum population difference among these SWCNTs can reach $\exp(-\Delta E_f/k_bT) \sim \exp(-30) \sim 10^{-14}$. The huge factor clearly indicates the great potential of chirality-selection in SWCNT growth. For example, the (10,0) zigzag SWCNT shows exceptional stability and its population in the sample can be estimated to be greater than 99%.

Despite the above analysis on the potential of achieving chirality-selected SWCNT growth, it has never been observed in experiments. In contrast, the as-grown SWCNT samples produced by experiments normally contain SWCNTs of various chiralities from AC to ZZ [9, 16, 17].

Theoretically, the atomistic simulations (including both molecular dynamic (MD) and Monte Carlo (MC) simulations) based on classical, semi-empirical or DFT potential energy surface (PES) have been extensively used to simulate the nucleation and growth of SWCNT for a long term [18-27]. Although great efforts have been paid, perfect SWCNT with identified chirality has never been achieved before and, consequently, the investigation to the SWCNT's chirality control through atomic simulation is prohibited.

Recently, we have developed a new generation of empirical PES for the C-Ni system based on the previous works [28, 29]with the parameters determining the C-Ni interactions in different coordinates fitted carefully based on many benchmark energies calculated by DFT. The C-C interactions and Ni-Ni interactions are described by the second-generation reactive empirical bond order (REBO) potential [30] and Sutton-Chen potential [31], respectively. A basin-hopping defect healing strategy is introduced into the molecular dynamic (MD) simulation. The newton's equation of the atom's motion is integrated by the Velocity-Verlet algorithm with the Berendsen thermostat [32] to maintain the temperature at 1,300 K during the MD simulation, which is closed to the typical experimental condition [see more details of the MD in ref. [15]]. Based on the above efforts, we are able to simulate perfect SWCNTs with identified chiralities [Fig. 2(a) and Fig. S4 in ref. [15]]. Figure 2(b, c) shows both the diameter and chiral angle distributions of the simulated SWCNTs on the $Ni_{32}$ catalyst particle. All the simulated SWCNTs grow in a tangential mode [33], and there exists a clear diameter selection [Fig. 2(b)]. Meanwhile, the statistics over chiral angle shows that SWCNTs with same diameter but different chiral angles have the same probabilities of being nucleated or there is no chiral angle selection. In another simulation, we further biased the energy difference between AC and ZZ SWCNTs to be ~ 2.0 eV/tube [see Fig. S5 in ref. [15]], but same distribution is resulted with no chiral angle selection [Fig. 2(c,e)].

Both the experiments and atomic simulations show an even chiral angle distribution of the SWCNTs, which drastically contradicts the *quasi-thermal-equilibrium* process in Eq. (1) [7]. Such puzzle can only be solved by one of the two routes: (i) there is actually no systematic IFE difference among SWCNTs of different chiral angles, or (ii) the nucleation of SWCNT can not be described by Eq. (1), which implies that the nucleation of SWCNT is a process that is far from thermal equilibrium. The DFT calculations shown in Fig. 1(j) and in previous studies [7, 11] confirmed that the IFE does depend on the edge structure or the chirality of the SWCNT. Therefore, the only route to solve the contradiction should be that the nucleation of SWCNT is a kinetic process, instead of *quasi-thermal-equilibrium*.

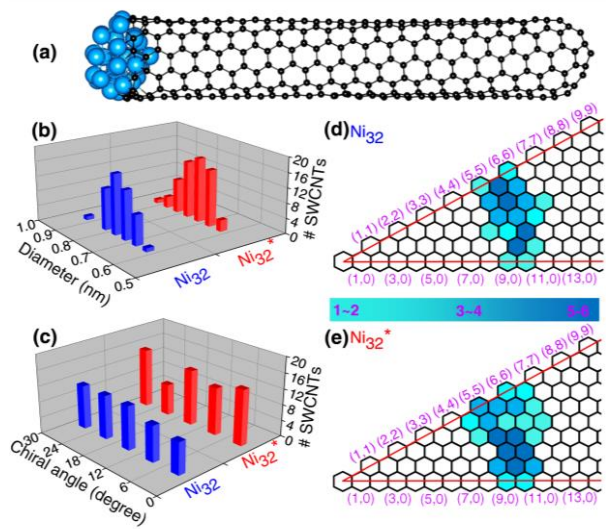

**FIG. 2** (color online). Distribution of the diameter and chiral angle of the grown SWCNTs by the atomic simulations. (a) A simulated defect-free SWCNT. (b-c) The diameter and chiral angle distributions of the simulated SWCNTs on a liquid $Ni_{32}$ catalyst particle with two different potential energy surfaces (PESs), respectively. The results obtained with normal C-Ni PES are labeled with $Ni_{32}$, while those simulated with a biased PES are labeled by $Ni_{32}^*$. (d-e) The population maps of the SWCNTs simulated by both the two PESs.

To achieve a deep insight into the kinetics of the SWCNT's chirality assignment in the catalytic growth, we recall the birth of a SWCNT. The nucleation of a SWCNT starts from the aggregation of carbon atoms to form small $sp^2$ carbon network on a catalyst particle surface [26, 27]. After that, guided by the curved catalyst particle surface, the network grows into a graphitic cap by adsorbing carbon atoms to form polygons, such as hexagons and pentagons. Without considering other defects (e.g., heptagons, octagons, $sp^3$ or dangling atoms), once the number of the pentagons in the graphitic cap reaches six, the cap becomes a mature hemisphere, which can be considered as a SWCNT infant because its chirality is uniquely determined by the relative positions of the six pentagons. In the following growth process, the addition of more carbon atoms into the SWCNT infant leads to the elongation of the SWCNT, without altering on the chirality because of the efficient healing of defects [9, 34].

To understand how the arrangement of the 6 pentagons in the cap determines a SWCNT's chirality, a smooth edged cap with five pentagons is considered [Fig. 3(a) and the similar results with other caps are shown in [15]]. For such a cap, incorporating one more pentagon is required to turn it into a SWCNT with well-defined chirality. In the cap shown in Fig. 3(a), there are 11 options of incorporating the $6^{th}$ pentagon during the addition of a new polygon ring [Fig. 3(b1-b11)], which turn the cap into 11 different SWCNTs, which are (11,0), right handed and left handed (10,1), (9,2), (8,3), (7,4) and (6,5) SWCNTs, respectively [Fig. 3(c1-c11)]. These chiralities cover all the population of the (m,n) SWCNT family of n + m = 11. If the new polygon ring composes all hexagons, the formation of the $6^{th}$ pentagon during the next polygon ring turns the cap into an arbitrary SWCNT of the n + m = 12 family such as the (6,6) armchair SWCNT, as shown in Fig. 3(d→e). Similarly, adding the $6^{th}$ pentagon far away from the cap center leads to the formation of a large SWCNT with any arbitrary chiral angle. This analysis clearly indicates that (*i*) the position of the $6^{th}$ pentagon fully controls the chirality of the SWCNT regardless the locations of the other five ones, and (*ii*) once the locations of $1^{st}$ to $5^{th}$ pentagon are determined, every SWCNT corresponds to a unique location of the $6^{th}$ pentagon. The above finding simplifies the origin of the SWCNT's chirality during nucleation, because only the $6^{th}$ or the last added pentagon matters. It's important to note that the $1^{st}$ to $5^{th}$ pentagons are normally formed at the central area of a cap. Their relative positions are fixed during further enlargement of the cap and thus can't be changed by the addition of the $6^{th}$ pentagon. The fixed locations of the $1^{st}$ to $5^{th}$ pentagons is a precondition of the aforementioned discussion, that the chirality is 100% controlled by the position of the $6^{th}$ pentagon.

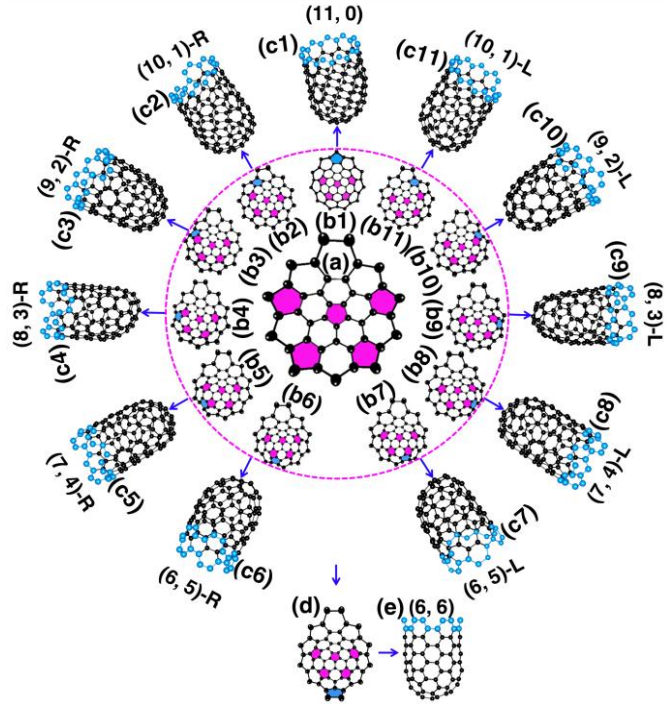

**FIG. 3** (color online). The nucleation of a SWCNT from a graphitic cap with five pentagons. (a) An immature graphitic cap with 5 pentagons. (b1-b11) The 11 options of adding the 6$^{th}$ pentagon into the cap when a new polygon ring is formed, and the resulted SWCNTs (c1-c11). (n,m)-L and (n,m)-R denote the left handed and right handed chiral SWCNTs, respectively. (d→e) The formation of the (6,6) SWCNT.

Based on above analysis, we propose that a SWCNT of any chiral angle has the same probability of being nucleated, if the addition of the 6$^{th}$ pentagon is not site selective. At the atomic level, as can be seen in Fig. 3(a)→(b1-b11), without considering the catalyst, the addition of the 6$^{th}$ pentagon in most sites occurs in a very similar manner--adding two carbon atoms onto two neighboring ZZ sites of the cap. For the VLS SWCNT growth, a liquid catalyst droplet has an isotropic surface, which ensures an identical environment around each site of the cap edge. Thus, in the VLS SWCNT growth, the probability of adding a pentagon onto any location of a cap edge is expected to be equal.

The above analysis indicates that the addition of the 6$^{th}$ pentagon during the SWCNT nucleation is a kinetic process, with equal probability at every potential site of the cap in the VLS growth. This successfully explains the confusing experimental observations that there is no chiral angle selection in most VLS SWCNT samples, although the IFEs between the SWCNTs and catalysts are very different. It is important to note that an isotropic surface of the liquid catalyst particle is assumed in drawing such a conclusion. Experimental observations have shown that the SWCNT growth on a solid catalyst particle via the vapor-solid-solid (VSS) mechanism is also possible [35, 36]. On a solid crystalline catalyst surface, there is a possibility that the sites along the circumference of a graphitic cap can be differentiated by the local environment of the anisotropic catalyst surface. In such circumstance, the formation of the 6$^{th}$ pentagon can be site-selective and, thus, chirality-selective nucleation of SWCNT could be achieved. As shown in Fig. 4, on an icosahedral $Ni_{55}$ surface, the addition of the 6$^{th}$ pentagon along the cap circumference leads to a systematic change of the formation energy of up to 2.0 eV. Among these sites, the two that correspond the (10,1) and (6,5) SWCNTs show exceptional stability near the ZZ and AC edges, respectively.

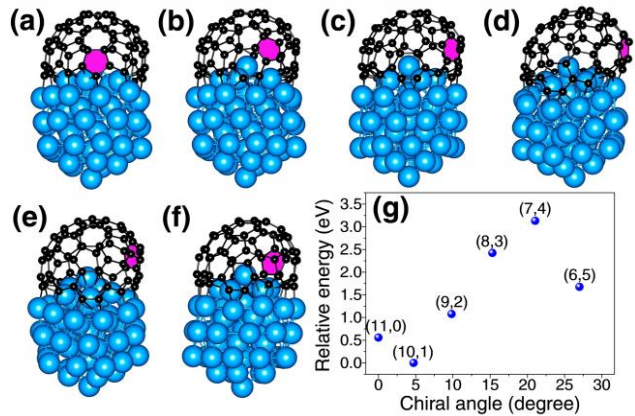

**FIG. 4** (color online). The relative formation energy of various matured graphitic caps (with six pentagons) on an icosahedral $Ni_{55}$ solid catalyst particle. (a-f) The caps which correspond SWCNTs of family n + m = 11, where the only difference of these caps is the location of the last pentagon (marked in purple). (g) The relative formation energies of the caps vs. the chiral angles of the corresponding SWCNTs.

Experimentally, the (6,5) SWCNTs have been synthesized with very large abundance in low temperature (~ 600-800 ℃) CVD growth by using Co or CoMo as catalyst (CoMoCat) [5, 6], in agreement with above theoretical analysis. Under such a low temperature, these catalysts may retain the solid crystalline structures. We also noticed that the chirality selection abruptly disappears at 850 $^{o}$C or higher temperature

[6]. Such a transition can be well explained by assuming that the melting point of the catalyst particles is between 800 and 850 $^{o}$C. Similarly, the preference growth of conducting SWCNT was also attributed to the catalyst formation with sharp edges, or, in another word, the catalyst must be in a crystalline structure [3].

The growth of high quality SWCNTs requires the high temperature (> 900 $^{o}$C) [37] and the most frequently used catalysts (Fe, Co and Ni) do not have high enough melting points to maintain the crystalline structure of the nano particles of a few nm in diameter. For this reason, there is normally no chrality selection in most SWCNT samples grown at high temperature. By simply examining the periodic table, the proper candidates can be Tantalum, Tungsten, Rhenium, Osmium (the four metal elements which melting point are about 3000 $^{o}$C) or their alloys with other elements. With this understanding, we propose that the design of high melting point catalysts to maintain the crystallinity of the catalyst particle as the route towards the direct growth of chirality-selected SWCNTs. Although a SWCNT tends to maintain it original chirality originated from the cap-to-SWCNT transition by the efficient healing of the topological defects [37], varying the growth condition may alter its chirality, for example, by varying the temperature slightly [see Fig. S7(a→c) in ref. [15]] during its growth [38]. For the change of SWCNT's chirality during SWCNT elongation, the IFE must play an important role. Based on this understanding, we propose another route of achieving chirality-selected SWCNT via alternating VLS and VSS growth by (i) Initiate the growth of SWCNTs at a certain temperature; (ii) varying the temperature to change the chirality of growing SWCNTs after a certain period; (iii) change the temperature back to the original; (iv) repeat (ii) and (iii) for many times [see Fig. S7 in ref. [15]].

In summary, through DFT calculations, a systematic change of the interfacial formation energy between SWCNT and catalyst was identified. The experimental puzzle of no chiral angle selection during SWCNT growth was successfully explained by the kinetic addition of $6^{th}$ pentagons into a graphitic cap during SWCNT nucleation. The previous experimental observations of the abundance of a few types of SWCNT synthesized at low temperature were also well explained by the solid catalyst particle.

Based on this understanding, two strategies of achieving chirality-selective growth in VLS and VSS growth are proposed—(*i*) using high melting catalyst (Ta, W, Re, Os) or their alloys as catalysts, and (*ii*) by changing the SWCNT's chirality during the elongation.

## Acknowledgements

The work of Hong Kong PolyU was supported by Hong Kong GRF research grant (G-YX4Q) and NSFC grant (21273189).The work of NKU is supported by the NSFC grant (21073097).

## References

[1] X. Tu, S. Manohar, A. Jagota, and M. Zheng, Nature **460**, 250 (2009).

[2] H. Liu, D. Nishide, T. Tanaka, and H. Kataura, Nat. Commun. **2**, 309 (2011).

[3] A. R. Harutyunyan *et al.*, Science **326**, 116 (2009).

[4] R. M. Sundaram, K. K. K. Koziol, and A. H. Windle, Adv. Mater. **23**, 5064 (2011).

[5] S. M. Bachilo, L. Balzano, J. E. Herrera, F. Pompeo, D. E. Resasco, and R. B. Weisman, J. Am. Chem. Soc. **125**, 11186 (2003).

[6] G. Lolli, L. Zhang, L. Balzano, N. Sakulchaicharoen, Y. Tan, and D. E. Resasco, J. Phys. Chem. B **110**, 2108 (2006).

[7] Y. Liu, A. Dobrinsky, and B. I. Yakobson, Phys. Rev. Lett. **105**, 235502 (2010).

[8] H. Wang, L. Wei, F. Ren, Q. Wang, L. D. Pfefferle, G. L. Haller, and Y. Chen, ACS Nano **7**, 614 (2013).

[9] F. Ding, A. R. Harutyunyan, and B. I. Yakobson, Proc. Nat. Acad. Sci. U.S.A. **106**, 2506 (2009).

[10] S. Reich, L. Li, and J. Robertson, Chem. Phys. Lett. **421**, 469 (2006).

[11] O. V. Yazyev, and A. Pasquarello, Phys. Rev. Lett. **100**, 156102 (2008).

[12] P. E. Blöchl, Phys. Rev. B **50**, 17953 (1994).

[13] G. Kresse, and J. Furthmüller, Phys. Rev. B **54**, 11169 (1996).

[14] J. P. Perdew, K. Burke, and M. Ernzerhof, Phys. Rev. Lett. **77**, 3865 (1996).

[15] See supplementary material at http://link.aps.org/supplemental/xxxx.

[16] S. M. Bachilo, M. S. Strano, C. Kittrell, R. H. Hauge, R. E. Smalley, and R. B. Weisman, Science **298**, 2361 (2002).

[17] K. Hirahara, M. Kociak, S. Bandow, T. Nakahira, K. Itoh, Y. Saito, and S. Iijima, Phys. Rev. B **73**, 195420 (2006).

[18] J. Gavillet, A. Loiseau, C. Journet, F. Willaime, F. Ducastelle, and J. C. Charlier, Phys. Rev. Lett. **87**, 275504 (2001).

[19] J.-Y. Raty, F. Gygi, and G. Galli, Phys. Rev. Lett. **95**, 096103 (2005).

[20] H. Amara, C. Bichara, and F. Ducastelle, Phys. Rev. Lett. **100**, 056105 (2008).

[21] E. C. Neyts, Y. Shibuta, A. C. T. van Duin, and A. Bogaerts, ACS Nano **4**, 6665 (2010).

[22] A. J. Page, H. Yamane, Y. Ohta, S. Irle, and K. Morokuma, J. Am. Chem. Soc. **132**, 15699 (2010).

[23] Y. Shibuta, and S. Maruyama, Chem. Phys. Lett. **437**, 218 (2007).

[24] J. Zhao, A. Martinez-Limia, and P. B. Balbuena, Nanotechnology **16**, S575 (2005).

[25] M. A. Ribas, F. Ding, P. B. Balbuena, and B. I. Yakobson, J. Chem. Phys. **131**, 224501 (2009).

[26] F. Ding, K. Bolton, and A. Rosén, J. Phys. Chem. B **108**, 17369 (2004).

[27] F. Ding, A. Rosen, and K. Bolton, J. Chem. Phys. **121**, 2775 (2004).

[28] Y. Yamaguchi, and S. Maruyama, Eur. Phys. J. D **9**, 385 (1999).

[29] A. Martinez-Limia, J. Zhao, and P. B. Balbuena, J. Mol. Model. **13**, 595 (2007).

[30] D. W. Brenner, O. A. Shenderova, J. A. Harrison, S. J. Stuart, B. Ni, and S. B. Sinnott, J. Phys.: Condens. Matter **14**, 783 (2002).

[31] A. P. Sutton, and J. Chen, Philos. Mag. Lett. **61**, 139 (1990).

[32] H. J. C. Berendsen, J. P. M. Postma, W. F. van Gunsteren, A. DiNola, and J. R. Haak, J. Chem. Phys. **81**, 3684 (1984).

[33] M. F. C. Fiawoo, A. M. Bonnot, H. Amara, C. Bichara, J. Thibault-Pénisson, and A. Loiseau, Phys. Rev. Lett. **108**, 195503 (2012).

[34] Q. Yuan, H. Hu, and F. Ding, Phys. Rev. Lett. **107**, 156101 (2011).

[35] S. Helveg, C. Lopez-Cartes, J. Sehested, P. L. Hansen, B. S. Clausen, J. R. Rostrup-Nielsen, F. Abild-Pedersen, and J. K. Norskov, Nature **427**, 426 (2004).

[36] S. Hofmann *et al.*, Nano Lett. **7**, 602 (2007).

[37] Q. Yuan, Z. Xu, B. I. Yakobson, and F. Ding, Phys. Rev. Lett. **108**, 245505 (2012).

[38] Y. Yao, Q. Li, J. Zhang, R. Liu, L. Jiao, Y. T. Zhu, and Z. Liu, Nat. Mater. **6**, 283 (2007).

Supplemental Material

# The Kinetics of Chirality Assignment in Catalytic Single-Walled Carbon Nanotube Growth

Ziwei Xu,[1] Tianying Yan,[2] Feng Ding[2,*]

[1]*Institute of Textiles and Clothing, Hong Kong Polytechnic University, Hong Kong, Peoples Republic of China*

[2]*Institute of New Energy Material Chemistry, College of Chemistry, Nankai University, Tianjin 300071, Peoples Republic of China*

---

[*]Corresponding author : feng.ding@polyu.edu.hk

# S-1: Methods

#1. Details of density functional theory (DFT) calculation

DFT calculations are performed with the VASP (Vienna Ab-initio Simulation Package) [1] [2]. The generalized gradient approximation (GGA) is adopted for the exchange correlation by using Perdew-Burke-Ernzerhof (PBE) functional, with the spin polarization taken into account [3]. The plan wave cutoff energy is set to be 400 eV and the projector-augmented wave (PAW) is used as the pseudopotential [4]. The convergence criterion for energy and force is set to be $10^{-4}$ eV and 0.01 eV/Å, respectively.

#2. The multi-scaled simulation of single-walled carbon nanotube (SWCNT) growth

The normal experimental duration of the SWCNT growth with the chemical vapor deposition (CVD) method is in the order of minutes to hours, which far exceeds the time scale that can be reached by the current molecular dynamic (MD) simulation. The rate limiting step in SWCNT growth is the defect-healing, which imposes high energy barrier of 1.5-3.0 eV, and such rare event is hardly observed in the MD simulation because the occurrence of each event costs many microseconds (μs) [5].

In order to solve such a time scaled issue, we developed a hybridized molecular dynamics − basin hopping (MD-BH) scheme to simulate the growth of the perfect SWCNTs. In such multi-scaled scheme, the addition and diffusion of carbon atoms on the catalyst surface as well as their bonding onto the SWCNT are simulated by the MD process, and the defects in the tube wall (i.e., pentagons and heptagons) formed by the addition of the carbon atoms can be healed by a BH process. In the BH process, a randomly selected C-C bond in the formed $sp^2$ carbon network is rotated by 90 degree and then relaxed to a local minimum. The energy difference between the final structure and the initial one, $\Delta E$, is used to determine the probability of acceptance by

$$p = \begin{cases} 1.0 & \Delta E < 0 \\ \exp(-\Delta E / k_b T) & \Delta E \geq 0 \end{cases} \qquad (S1)$$

where $T$ is the temperature of simulation and the $k_b$ is the Boltzmann constant.

MD simulation starts with a bare nickel cluster, $Ni_{32}$, with a carbon atom randomly added onto the surface every 30 ps to mimic the catalytic feedstock decomposition in CVD experiments. The Berendsen thermostat [6] is used to maintain the SWCNT growth temperature at 1,300 K, which is closed to the typical experimental condition. For every trajectory, MD simulation is firstly performed for 60 ps (with the addition of two C atoms) and then BH process is performed to heal the potential defects formed during the MD simulation. By carefully control the parameters of the simulation, perfect SWCNTs can be achieved.

## S-2: Definition of interfacial formation energy (IFE)

The IFE ($E_f$) of a SWCNTs on a liquid $Ni_{55}$ is

$$E_f = E_{FE} - E_b \qquad (S2)$$

in which $E_{FE}$ and $E_b$ are the formation energy of the free SWCNT end and the SWCNT-metal binding energy, respectively, and $E_b$ is

$$E_b = E_{NT} + E_{Ni} - E_{NT@Ni} \qquad (S3)$$

in which $E_{NT@Ni}$ is the energy of SWCNT attached on $Ni_{55}$, $E_{NT}$ and $E_{Ni}$ are energies of the isolated SWCNT and $Ni_{55}$, respectively. $E_{FE}$ in Eq. (S2) is

$$E_{FE} = 0.5*(2*E_{NT2} - E_{NT1}) \qquad (S4)$$

in which $E_{NT1}$ is the energy of a longer SWCNT and $E_{NT2}$ is the energy of a shorter SWCNT, which is obtained by cutting the longer SWCNT into two equal segments (see Fig. S1). The factor 0.5 refers the fact that two open ends are formed when a SWCNT is cut into two SWCNTs (see Fig. S1).

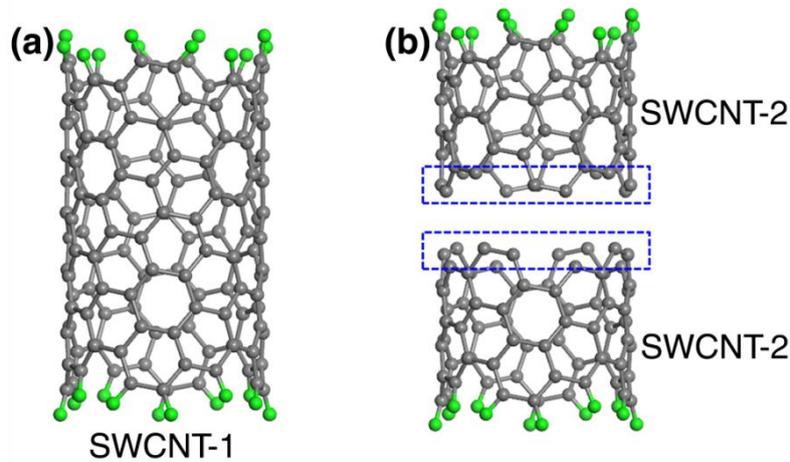

**FIG. S1** Schematic diagram of (a) SWCNT-1 and (b) SWCNT-2 for the calculation of the formation energy $E_{FE}$ of the free SWCNT end. The green atoms are hydrogen atoms. Two open

ends (blue dashed rectangular) are formed by cutting the longer SWCNT-1 into two shorter SWCNT-2 segments. The $E_{FE}$ is defined as Eq. (S4).

## S-3. Liquid Ni$_{55}$ particle formation

To obtain a liquid Ni$_{55}$ particle, MD simulation with Sutton-Chen potential [7] is performed at 1,500 K, which is greatly above the melting point of the cluster, for 10 ns. Randomly selected liquid Ni$_{55}$ structures in the MD simulation were used to represent the liquid catalyst particles in the IFE calculation (see Fig. S2).

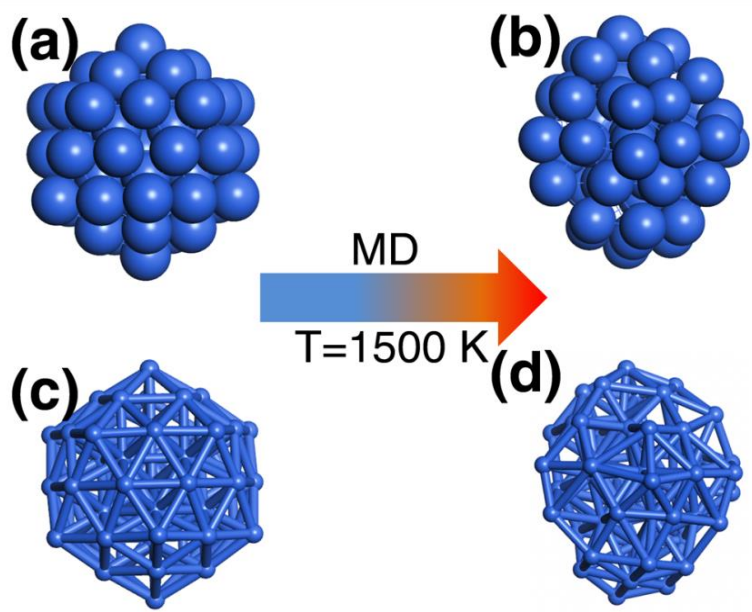

**FIG. S2** An icosahedral cluster of Ni$_{55}$ is melted at 1,500 K and liquid particles are obtained during the MD simulation. (a→b) CPK model and (c→d) ball-stick model of the solid and liquid Ni$_{55}$ particles.

## S-4: The IFE calculation

To achieve a reasonable IFE for each SWCNT attached to the liquid Ni$_{55}$ catalyst particle, six different melting Ni$_{55}$ structures are used for DFT calculations and thus six IFEs are obtained for each SWCNT as shown in Fig. S3. Among the six IFEs, the smallest one is used to represent the IFE of a SWCNT-catalyst interface as plotted in Fig. 1(j) of the main text considering the high probability of it being formed during

SWCNT growth. It's worth to note that the statistics of the data and the calculated IFEs of SWCNTs attached a solid icosahedral catalyst particle represent same trend (see Fig. S3).

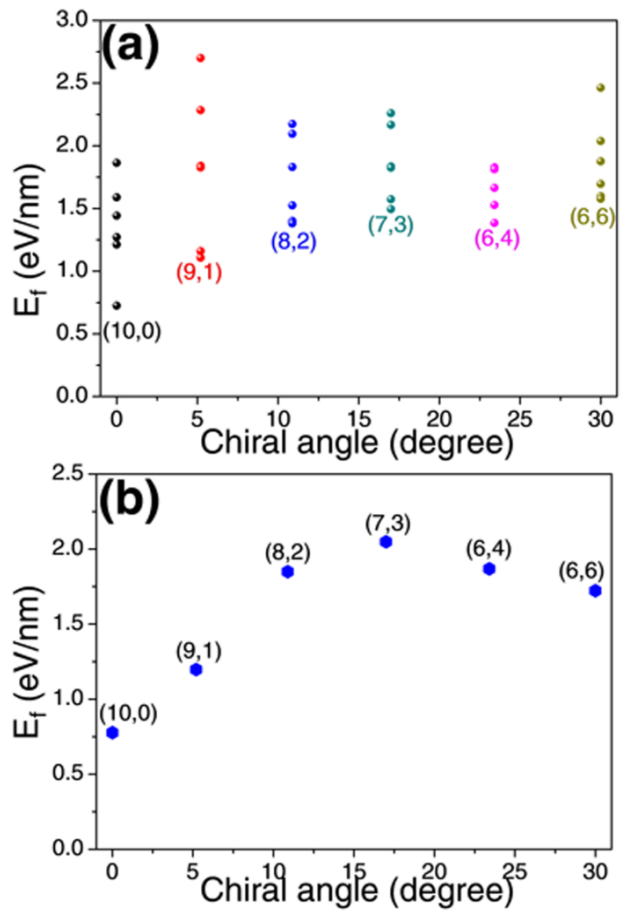

**FIG. S3** IFEs calculated by attaching SWCNTs onto liquid and solid $Ni_{55}$ particles. (a) The six interfacial formation energies (IFEs) calculated by attaching each SWCNT onto six different liquid catalyst particles with the DFT method (see the methods). The detailed values are listed in table S1. (b) IFEs calculated by attaching SWCNTs onto solid catalyst particle (icosahedral $Ni_{55}$ cluster).

**Table S1.** The interfacial energies (IFEs in eV/nm) calculated by attaching SWCNT onto liquid Ni$_{55}$ particles with the DFT method. #1 - #6 refers six liquid Ni$_{55}$ particles. The lowest IFE for each SWCNT is marked in bold.

| IFE / CNT | #1 | #2 | #3 | #4 | #5 | #6 |
|---|---|---|---|---|---|---|
| (10,0) | 1.8637 | 1.5896 | 1.2123 | 1.2719 | 1.4431 | **0.7245** |
| (9,1) | 1.8258 | 2.2845 | **1.1074** | 1.1591 | 2.7004 | 1.8410 |
| (8,2) | 1.8294 | 1.4007 | 1.5244 | **1.3816** | 2.1737 | 2.0951 |
| (7,3) | **1.4950** | 1.5729 | 2.2597 | 2.1671 | 1.8351 | 1.8242 |
| (6,4) | 1.828 | 1.8118 | 1.5276 | **1.3847** | 1.6636 | 1.8160 |
| (6,6) | 1.6959 | 2.4621 | 2.0387 | **1.5764** | 1.8761 | 1.5971 |

## S-5: Simulated SWCNTs

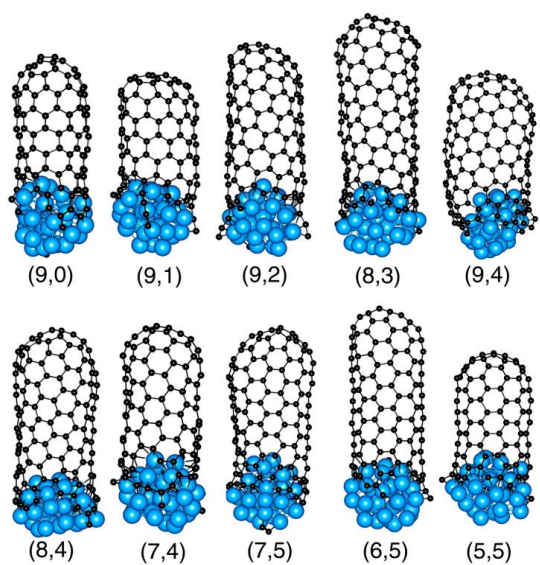

**FIG. S4** Some simulated SWCNTs obtained by the hybrid MD-BH scheme.

# S-6: Models used for the interfacial formation energy (IFE) calculation.

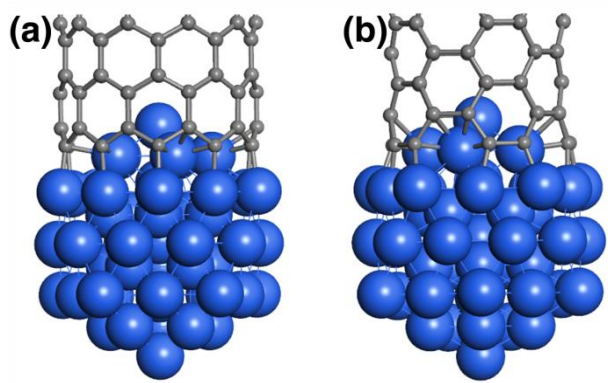

**FIG. S5** (a) (10,0) zigzag (ZZ) SWCNT attached to the $Ni_{55}$ particle. (b) (5,5) armchair (AC) SWCNT attached to the $Ni_{55}$ particle. By changing the parameters of the carbon-metal potential, the IFE of ZZ SWCNT@$Ni_{55}$ can be tuned to be 2.0 eV/nm higher than that of the AC SWCNT@$Ni_{55}$.

# S-7: The random chirality assignment from other caps with five pentagons.

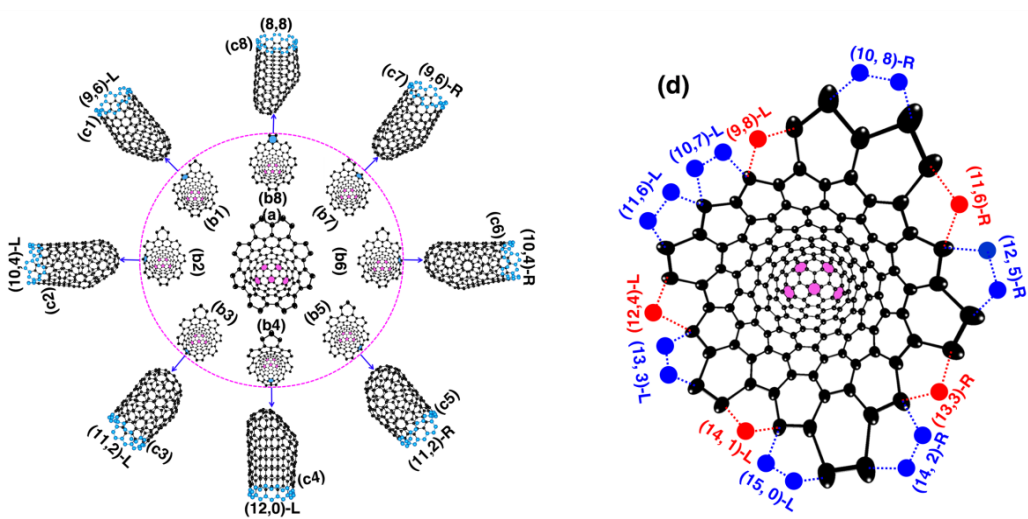

**FIG. S6** (a) An graphitic cap with seven AC and one ZZ sites on the edge. (b1-b8) the eight options of forming the 6$^{th}$ pentagon into the cap and the resulted SWCNTs (c1-c8). (n,m)-L and (n,m)-R denote the left handed and right handed chiral SWCNTs, respectively. (d) The chiral indexes of the resulted SWCNTs by adding the 6$^{th}$ pentagon onto different edge sites of the cap which is mixed with AC and ZZ sites. It can be clearly seen that the resulted SWCNTs have randomly assigned chiralities either the cap edge is dominated by AC sites or mixed with AC and ZZ sites.

## S-8: Illustration of the strategy of achieving chirality-selected SWCNT growth by varying the temperature during SWCNT elongation.

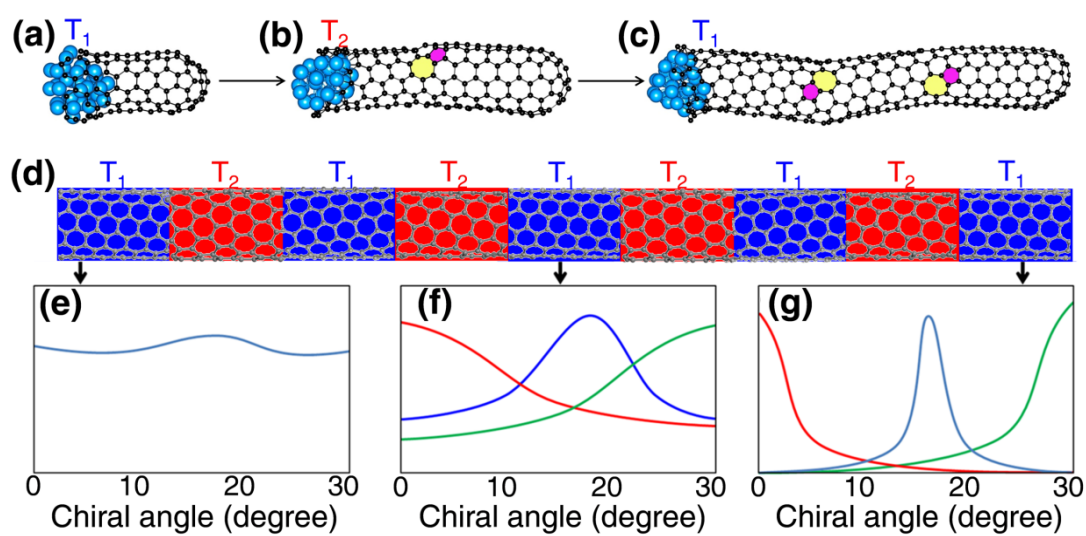

**FIG. S7** (a→c) The slight change of the temperature during the SWCNT elongation can lead to the change of the SWCNT chirality. Repeated change of the growth temperature (d) can drive the even chiral-angle distribution of initial SWCNTs to the final narrowed distribution with the most stable interfacial formation energy (e→g).

## Supplemental References

[1] P. E. Blöchl, Phys. Rev. B **50**, 17953 (1994).
[2] G. Kresse, and J. Furthmüller, Phys. Rev. B **54**, 11169 (1996).
[3] J. P. Perdew, K. Burke, and M. Ernzerhof, Phys. Rev. Lett. **77**, 3865 (1996).
[4] G. Kresse, and D. Joubert, Phys. Rev. B **59**, 1758 (1999).
[5] Q. Yuan, Z. Xu, B. I. Yakobson, and F. Ding, Phys. Rev. Lett. **108**, 245505 (2012).
[6] H. J. C. Berendsen, J. P. M. Postma, W. F. van Gunsteren, A. DiNola, and J. R. Haak, J. Chem. Phys. **81**, 3684 (1984).
[7] A. P. Sutton, and J. Chen, Philos. Mag. Lett. **61**, 139 (1990).